\documentclass[11pt,letterpaper]{article}

\usepackage[margin=1in,headheight=14pt]{geometry}
\usepackage[T1]{fontenc}
\usepackage[utf8]{inputenc}
\usepackage{textcomp}
\usepackage{newtxtext,newtxmath}
\usepackage{microtype}
\usepackage{indentfirst}
\usepackage{xcolor}
\usepackage{enumitem}
\usepackage{titlesec}
\usepackage{fancyhdr}
\usepackage{hyperref}
\usepackage{xurl}
\usepackage{bookmark}

\definecolor{BMNavy}{HTML}{17324D}
\definecolor{BMLink}{HTML}{1F5A7A}

\hypersetup{
  pdftitle={The Burst Market: The Next Leap for Humanity},
  pdfauthor={Vincent Yuansang Zha},
  pdfsubject={A conceptual model for brief, paid, immediate human services},
  pdfkeywords={Burst Market, minimum viable exchange, transaction costs, generative market infrastructure, human-capability utilization gap, artificial intelligence and employment},
  colorlinks=true,
  linkcolor=BMNavy,
  urlcolor=BMLink,
  citecolor=BMNavy,
  pdfcreator={LaTeX}
}
\titleformat{\section}
  {\Large\bfseries\color{BMNavy}}
  {\thesection}{0.65em}{}
\titleformat{\subsection}
  {\large\bfseries\color{BMNavy}}
  {\thesubsection}{0.65em}{}
\titlespacing*{\section}{0pt}{2.2ex plus 0.6ex minus 0.2ex}{0.8ex}
\titlespacing*{\subsection}{0pt}{1.7ex plus 0.4ex minus 0.2ex}{0.55ex}

\setlist{leftmargin=2.1em,itemsep=0.25em,topsep=0.45em}

\fancypagestyle{plain}{%
  \fancyhf{}
  \fancyhead[L]{\small\itshape The Burst Market}
  \fancyhead[R]{\small Vincent Yuansang Zha}
  \fancyfoot[C]{\small\thepage}
  
}
\fancypagestyle{firstpage}{%
  \fancyhf{}
  \fancyfoot[C]{\small\thepage}
  
}

\newenvironment{bmproposition}[1]{%
  \par\medskip
  \noindent\begin{minipage}{\textwidth}
  \color{BMNavy}\rule{\textwidth}{0.6pt}\par\smallskip
  \noindent\textbf{#1}\par\smallskip
  \color{black}\itshape
}{%
  \par\smallskip\color{BMNavy}\rule{\textwidth}{0.35pt}\color{black}
  \end{minipage}\par\medskip
}

\title{\vspace{-1.5em}\textcolor{BMNavy}{\bfseries The Burst Market:\\[0.18em]The Next Leap for Humanity}}
\author{Vincent Yuansang Zha\\[0.45em]
  \small ORCID: \href{https://orcid.org/0000-0002-8435-9111}{\nolinkurl{https://orcid.org/0000-0002-8435-9111}}}
\date{}

\begin{document}
\maketitle
\thispagestyle{firstpage}

\begin{abstract}

Many useful human contributions remain unexchanged because exchange frictions exceed their value. This conceptual paper proposes the Burst Market (BM): a market for brief, paid, immediate human services delivered through live video. BM seeks to lower the minimum viable exchange through PIVOT---Provider-set price, Immediate access, Video interaction, Open participation, and Time-metered payment. A four-part friction map covers buyer access, provider market access, capability packaging, and assurance and governance. BM identifies seven rebalances, including AI job disruption versus underused human capability, and serves both scarce and underused capacity. As generative market infrastructure, BM may let people create services that established occupations and platforms cannot foresee. Seven conceptual propositions organize its mechanisms and predictions. If these effects emerge at scale, BM could represent the next leap for humanity.
\end{abstract}

\vspace{0.75em}

\section{Introduction}\label{introduction}

Many useful human contributions are too small, informal, personal, or situational to become jobs, projects, appointments, courses, or conventional services. A person may know one useful fact about a workplace, understand an unusual repair, observe a distant place, offer an ordinary customer\textquotesingle s reaction, perform one personalized song, or simply be willing to talk. Someone else may value that contribution, yet no practical exchange occurs.

This paper calls the difference between potentially useful human contribution and what existing institutions can practically discover, elicit, exchange, and compensate the \emph{human-capability utilization gap}. The gap does not imply that every ability has a buyer or that leisure, family life, care, friendship, and voluntary activity are wasted. It identifies a narrower institutional problem: even when one person would willingly help and another would willingly pay, the exchange may still be too difficult to arrange.

Consider a three-minute need. A prospective tenant wants a resident to show whether an apartment is noisy. A designer wants an ordinary person\textquotesingle s reaction to a new interface. Someone facing an unfamiliar object wants an experienced owner to look at it. The interaction may take less time than finding the person, establishing trust, explaining the need, negotiating, scheduling, connecting, and paying. When this surrounding effort costs more than the help is worth, the exchange does not happen.

The problem is one of transaction granularity. Using a market has costs: people must search, compare, bargain, coordinate, monitor, and enforce (Coase, 1937; Williamson, 1981). Digital markets reduce some costs and change which transactions become practical (Malone, Yates, and Benjamin, 1987; Bakos, 1997), but most service and labor arrangements remain organized around units large enough to carry substantial fixed costs.

AI makes this neglected problem more urgent. Rapid advances in generative AI are transforming work and intensifying concern about employment disruption (Gmyrek et al., 2025). Society is investing extraordinary effort in expanding machine capability while leaving substantial human capability undiscovered and unused. Better infrastructure for exchanging human contribution is therefore both a longstanding opportunity and a response to AI-led employment change.

This paper asks:

\begin{quote}
How can a market reduce the exchange frictions that prevent brief human contributions from being traded, and how might that change the capabilities, services, and work that become economically visible?
\end{quote}

The proposed answer is the Burst Market (BM): a market for brief, paid, immediate human services delivered through live video. Providers set their rates and indicate current availability. A buyer can find a provider, connect immediately, and pay for actual connected time. Participation is open rather than confined to established occupations or recognized experts.

PIVOT summarizes this architecture: Provider-set price, Immediate access, Video interaction, Open participation, and Time-metered payment. Each element addresses a different obstacle. Together they seek to lower the \emph{minimum viable exchange}: the smallest interaction for which both parties expect nonnegative net value. The buyer expects the benefit to exceed the price plus buyer-side exchange costs; the provider expects payment to exceed delivery, opportunity, and provider-side exchange costs.

BM offers one response to the AI-employment challenge. Rather than competing with AI only inside existing occupations, people could offer contributions that present labor markets overlook and experiment with new human-centered services. Their value may arise from firsthand experience, current physical access, human presence, identity, authority, accountability, participation, or live performance. Other forms may be impossible to predict before the market exists.

The paper makes four contributions. Terminologically, the initial 2021 version introduced \emph{Burst Market} for this market form (Zha, 2021); this paper introduces the terms \emph{generative market infrastructure} and \emph{human-capability utilization gap}. Theoretically, it connects minimum viable exchange, a four-part BM friction map, capability-packaging cost, interactive elicitation, two capacity conditions, and seven rebalances. Architecturally, it introduces PIVOT as a strict five-element framework. Historically, earlier systems implemented different subsets of PIVOT, but the review found no documented predecessor to BM that unambiguously combined all five elements.

The claims are conceptual predictions, not reported effects. If BM makes many presently impractical exchanges viable, small interactions could accumulate into wider access, new income, lower costs for suitable professional help, faster collaboration, and unforeseen services and industries. These possible ripple effects motivate the model, its conceptual propositions, and the research agenda.

\section{Academic Foundations and Prior Art}\label{academic-foundations-and-prior-art}

\subsection{Conceptual approach}\label{conceptual-approach}

This is a phenomenon-based conceptual paper. It synthesizes theory, defines a market form, identifies mechanisms, derives propositions, and sets an empirical agenda rather than estimating an observed effect (Jaakkola, 2020). The targeted review draws on transaction-cost economics, electronic markets, knowledge exchange, platforms, AI and labor, user innovation, professional services, and documented predecessor systems. The prior-art search examined scholarly literature, patent records, archived product materials, official records, and contemporaneous reporting for combinations of the five PIVOT elements.

\subsection{Transaction costs and small exchanges}\label{transaction-costs-and-small-exchanges}

Coase\textquotesingle s central insight was that using the price mechanism is itself costly. Finding relevant prices and negotiating separate exchanges can make market coordination less attractive than organization within a firm (Coase, 1937). Williamson (1981) made the transaction the unit of analysis and showed that uncertainty, repeated interaction, and relationship-specific investment affect appropriate governance. Dahlman (1979) grouped transaction costs into search and information, bargaining and decision, and policing and enforcement.

These costs weigh especially heavily on a small service. Searching or paying may require similar effort whether the service lasts two minutes or two hours, and fixed payment costs matter disproportionately at small values (ten Raa and Shestalova, 2004). The economic boundary is not a universal duration but the point below which direct value cannot carry the fixed cost of exchange.

Electronic markets can move this boundary. Information systems reduce coordination and comparison costs, potentially shifting activity from organizational hierarchy toward markets (Malone, Yates, and Benjamin, 1987). Searchable seller attributes and prices can improve matching (Bakos, 1997). The relevant question is whether BM leaves a lower total cost for a particular short service.

\subsection{Packaging and eliciting human capability}\label{packaging-and-eliciting-human-capability}

Possessing useful knowledge is not the same as being able to sell it. A provider may have to identify, organize, explain, demonstrate, and market what is useful. Producing a reusable article, course, or video can require far more time than consuming it. Creating reusable content is production; packaging capability for a particular trade is also an exchange cost.

Knowledge research has long distinguished tacit knowledge from knowledge that has been made explicit. Codification may require new models, language, and messages, and some knowledge remains uncodified because the expected benefit does not justify the effort (Cowan, 2001). Studies of personal knowledge exchanges also find that tacitness, situatedness, and complexity can impede the mobilization of expertise even when an online market exists (Gosain, 2007).

BM changes this burden through \emph{interactive elicitation}. The buyer supplies a concrete situation and helps structure the service through questions. Video lets the parties point, observe, demonstrate, test, and correct. The provider need not anticipate every user or produce a complete one-way explanation. The term has a history in statistical elicitation (Kadane et al., 1980); its BM significance is economic: interaction can lower the cost of packaging a small human capability for exchange.

This mechanism is especially important for ordinary providers. A person may be capable of answering a concrete question while being unable or unwilling to publish a polished tutorial, build an audience, or present a logically complete lecture. BM lets some value emerge in dialogue rather than requiring formal exposition before demand is known.

\subsection{Platforms, generativity, and YouTube}\label{platforms-generativity-and-youtube}

BM is a two-sided market: useful providers attract buyers, while demand attracts providers (Rysman, 2009). Effective markets need relevant thickness, manageable congestion, and safe participation (Roth, 2008). What matters is whether a suitable, trustworthy provider is visible and available when needed. More listings can reduce matching if they increase search burden without improving discovery (Li and Netessine, 2020).

Platforms can also enable decentralized participants to innovate. This paper introduces \emph{generative market infrastructure} as a market architecture that allows participants to discover and create service forms that the market designer has not predefined. The term builds on established research about generative digital infrastructure and user innovation rather than claiming that platform generativity itself is new (von Hippel, 2005; Acquatella, Fernandez, and Houy, 2022).

YouTube provides a useful analogy. By lowering entry and distribution barriers outside television and film, it helped ordinary participants create unforeseen genres, identities, audiences, and business models (Burgess and Green, 2018; Cunningham, Craig, and Silver, 2016). BM could extend this generativity from user-created content to user-created live services. A ten-minute public video may require research, scripting, recording, editing, promotion, and audience development. A burst service can be produced and consumed together, without durable content or a regular publishing commitment. YouTube supports the plausibility of unforeseen creation, not a prediction that BM will achieve comparable scale.

\subsection{Prior systems and scoped originality}\label{prior-systems-and-scoped-originality}

Brief paid human interaction has substantial prior art. Historical Keen offered provider-set per-minute rates and immediate telephone connection (Microsoft, 2001; Jacob, Faber, and Van der Linden, 2007). Today\textquotesingle s Keen is positioned around live spiritual advisors, and new customers typically receive five minutes for a fixed \$1 before the advisor\textquotesingle s rate applies (Keen, n.d.-a, n.d.-b). Ether let users charge for telephone or email services using provider-controlled per-minute or per-call pricing. BitWine combined directories, provider rates, and live Skype contact. Wengo developed a European market for remotely delivered knowledge and talent (Oswald, 2006; Gonzalez, 2006; ARCEP/IDATE, 2006). Collectively, these systems implemented several PIVOT elements but not the complete framework.

Google Helpouts moved closer to PIVOT. It offered broad live video help, instant or scheduled connection, pricing, ratings, and per-minute or per-session payment, but participation depended on selected categories, invitations, screening, recognized experts, and established brands (Manber, 2013; Ribeiro, 2013). \emph{Anyone} offered paid five-minute advice calls but used a fixed block (Lomas, 2021).

BM did not invent paid calls, expert markets, seller rates, live advice, time billing, or monetization of spare knowledge. Its originality lies in naming and theorizing the Burst Market, defining minimum viable exchange and the human-capability utilization gap, organizing the four-part BM friction map, and articulating PIVOT as a complete architecture. Earlier systems implemented different subsets of PIVOT, but the review found no documented predecessor to BM that unambiguously combined all five elements.

\subsection{AI, employment, and person-dependent value}\label{ai-employment-and-person-dependent-value}

Automation displaces labor in some tasks while productivity, reorganization, and new tasks can create demand elsewhere (Acemoglu and Restrepo, 2019). Generative AI has increased productivity in some workplaces, with effects varying by task and worker experience (Brynjolfsson, Li, and Raymond, 2025). BM does not depend on a permanent boundary around what AI cannot do. Its claim is institutional: as tasks change, people need better ways to offer smaller and different contributions. Human participation can itself be valuable when a buyer wants firsthand experience, current access, actual human presence, a particular identity, valid consent, accountable judgment, ordinary participation, or personalized live performance. Even when AI-generated empathy is rated highly, people may still prefer a human source (Rubin et al., 2025; Wenger, Cameron, and Inzlicht, 2026).

\section{The Burst Market Model}\label{the-burst-market-model}

\subsection{Definition, unit, and boundary}\label{definition-unit-and-boundary}

A Burst Market is a market for brief, paid, immediate human services delivered through live video. In this open digital market, ordinary people and professionals set their time rates, indicate immediate availability, and receive payment for actual connected duration.

BM is a market form, not a particular company. A \emph{burst service} is the offered human contribution. A \emph{burst interaction} is its live delivery. A \emph{burst transaction} is the purchase and sale. A platform supplies the infrastructure for discovery, connection, payment, and governance.

``Burst'' is an economic property rather than a fixed duration. A service is burst-length when its useful core is brief and the ordinary cost of arranging it would otherwise be large relative to the contribution. Seconds and minutes are the clearest cases. A longer conversation may also qualify when it requires little preparation and ends when the need is met.

Knowledge and advice are central but do not define the market. A contribution can also be observation, experience, participation, judgment, presence, identity-specific access, conversation, feedback, or performance.

\subsection{The four-part BM friction map}\label{the-four-part-bm-friction-map}

This paper uses a four-part BM friction map to organize the transaction costs and other exchange frictions surrounding a burst exchange.

\begin{enumerate}
\def\labelenumi{\arabic{enumi}.}
\item
  \textbf{Buyer access costs} include recognizing that a provider exists, searching, comparing identity and fit, checking price and availability, describing the need, connecting, and paying.
\item
  \textbf{Provider market-access costs} include making a capability visible, describing an offer, finding customers, establishing credibility, negotiating, scheduling, collecting payment, and waiting for demand.
\item
  \textbf{Capability-packaging costs} include identifying what can be offered and turning fragmented, tacit, or situational capability into a service the buyer can understand and use.
\item
  \textbf{Assurance and governance costs} include identity, reputation, credentials, consent, privacy, safety, records, payment protection, and dispute resolution.
\end{enumerate}

The first two groups show why both sides matter: buyer access may be easy while provider acquisition and waiting remain costly. The third explains why a person can possess value without being able to publish or professionalize it. The fourth includes the safeguards that enable trustworthy exchange.

The \emph{minimum viable exchange} is the smallest interaction for which both parties expect nonnegative net value. The buyer expects the benefit to exceed the price plus buyer-side exchange costs; the provider expects payment to exceed delivery, opportunity, and provider-side exchange costs. Reducing the price alone may not make an interaction viable if finding or trusting the provider remains difficult. Conversely, excellent matching does not help if the provider must spend an hour packaging a two-minute contribution.

\begin{bmproposition}{Proposition 1 --- Lower friction makes smaller exchanges possible}
When exchange frictions that are largely independent of service duration fall for both parties, some human contributions that were previously too small to trade become viable.
\end{bmproposition}

\subsection{Capability packaging and interactive elicitation}\label{capability-packaging-and-interactive-elicitation}

Traditional content and service markets often require a seller to define the product before meeting the buyer. This favors people who can write, teach, present, market, or create media and overlooks knowledge that is practical, narrow, contextual, or difficult to explain without seeing the problem.

Interactive elicitation divides the work of specification. The buyer supplies the actual need; the provider asks or answers questions; video conveys objects, facial expression, gesture, and environment; and each response determines what comes next. The exchange does not need a complete script or predefined deliverable.

Someone who has repaired an unusual appliance may be unable to write a general manual. In a video interaction, the buyer can show the sound, part, and failed step, and the provider can explain only what becomes relevant. Practical capability becomes offerable without first becoming a formal course.

\begin{bmproposition}{Proposition 2 --- Interactive elicitation reduces packaging cost}
When the buyer can structure the service through questions and visual context, a provider need not package a complete one-way product before contributing.
\end{bmproposition}

\subsection{PIVOT}\label{pivot}

PIVOT defines the strict BM architecture.

\textbf{P --- Provider-set price.} Each provider decides the time rate at which an interruption or contribution becomes worth offering. A visible rate reduces repeated negotiation and recognizes differences in capability, opportunity cost, demand, identity, and willingness to participate.

\textbf{I --- Immediate access.} A buyer sees who is available and can begin without arranging an appointment, reducing scheduling, waiting, and loss of context.

\textbf{V --- Video interaction.} Live video removes distance while preserving dialogue, gesture, demonstration, and observation of the buyer\textquotesingle s or provider\textquotesingle s setting.

\textbf{O --- Open participation.} The market is not limited to recognized experts or predefined occupations. An ordinary person may offer a reaction, experience, observation, conversation, small talent, or other contribution. Openness applies to the market as a whole; particular services may still require relevant qualifications.

\textbf{T --- Time-metered payment.} The buyer pays for actual connected duration rather than purchasing a full appointment, session, or project. Metering lowers the minimum commitment.

\begin{bmproposition}{Proposition 3 --- PIVOT elements are hypothesized to be complementary}
The five PIVOT elements define the burst form and are hypothesized to be complementary for brief, unplanned, video-suitable services. A two-minute video answer may remain impractical if the buyer must book an appointment or pay for a full hour. The full configuration combines immediate discovery with provider-set pricing, video, open entry, and time metering.
\end{bmproposition}

\subsection{Both scarce and underused capacity}\label{both-scarce-and-underused-capacity}

BM can serve two conditions that look opposite.

Some people receive more requests than they can answer. They can choose when to be available and raise their price until calls become manageable. The rate compensates interruption and directs buyers who do not require that particular person toward alternatives.

Others have time or useful abilities that existing markets do not know how to buy: ten free minutes, an ordinary perspective, a secondary skill, local access, or one small performance. BM provides a place to offer the contribution.

These are not permanent classes. The same person may be highly demanded for one capability and underused for another. Time also has valuable nonmarket uses (Becker, 1965).

\begin{bmproposition}{Proposition 4 --- BM serves both scarce and underused capacity, subject to demand}
BM can help an over-requested person make limited access manageable and help another person offer a small capability that existing markets overlook.
\end{bmproposition}

\subsection{Generative market infrastructure}\label{generative-market-infrastructure}

Most platforms begin with known categories. BM can also let providers describe what they are willing to do and let buyers reveal needs that have no established label. Because entry, delivery, and settlement are small, unsuccessful experiments need little commitment. Successful patterns can become recognizable service categories, occupations, or larger businesses.

This is the sense in which BM is generative market infrastructure. It does not merely distribute a catalog of known services more efficiently. It supplies conditions in which participants can create and test the catalog itself.

\begin{bmproposition}{Proposition 5 --- Open exchange encourages service innovation}
Lowering the cost of offering and testing a burst service should increase experimentation with services not represented in established occupations or platform categories. The most consequential categories may be impossible to name before people begin using the market.
\end{bmproposition}

\section{Seven Rebalances}\label{seven-rebalances}

BM may rebalance seven mismatches in present labor, service, and communication institutions. These rebalances connect the market architecture to wider social effects.

\subsection{AI job disruption vs. underused human capability}\label{ai-job-disruption-vs-underused-human-capability}

AI threatens established jobs while much human capability remains unused. Society is rapidly expanding machine capability but lacks comparable infrastructure for discovering small, current, personal, and unconventional human contributions. BM opens more of that capability to exchange and allows new human-centered services to form outside existing occupations.

\begin{bmproposition}{Proposition 6 --- BM provides an employment response to AI job disruption}
As AI disrupts existing work, BM lowers the barrier to earning from human contribution and permits new human-centered services to emerge. It can support a shift from some conventional employment and unpaid contribution into burst exchange, subject to demand and market development.
\end{bmproposition}

\subsection{Burst service needs vs. bundled work}\label{burst-service-needs-vs-bundled-work}

Many needs are brief, but service and labor markets package supply into jobs, projects, appointments, sessions, and contracts. A useful three-minute contribution can remain unavailable because it cannot support a one-hour appointment. BM brings the purchased unit closer to the actual need.

\subsection{Immediate needs vs. delayed service access}\label{immediate-needs-vs-delayed-service-access}

A need may be most valuable now, while conventional services require searches, messages, quotations, intake, and appointments. By displaying current availability and enabling immediate connection, BM reduces the gap between recognizing a need and receiving help.

\subsection{Lifelong learning vs. limited working}\label{lifelong-learning-vs-limited-working}

People learn throughout life, but opportunities to earn and contribute are concentrated in conventional working years and schedules. Contzen et al. (2017) use ``lifelong working'' for extended working lives and work beyond retirement. BM gives \emph{Lifelong Working} a voluntary, granular meaning: the option to contribute and earn throughout life, in amounts and at times chosen by the individual. Older people can share accumulated experience, while students, caregivers, and people unable to sustain conventional employment can participate around other needs.

\subsection{Service use vs. provider compensation}\label{service-use-vs-provider-compensation}

Professional intake illustrates this imbalance. When a prospective client receives useful preliminary attention without paying, the provider must absorb its cost or recover it elsewhere. Legal services already carry substantial business-model and access costs (Hadfield, 2019), while common fee structures can fit incidental needs poorly (Felső, Onderstal, and Seldeslachts, 2022). BM can make suitable preliminary time independently payable, aligning compensation with the minutes actually used while leaving deliberate pro bono or free assistance intact.

\subsection{Easy spending vs. difficult earning}\label{easy-spending-vs-difficult-earning}

People can spend money in countless immediate ways, yet earning generally requires a job, contract, business, audience, or long production process. BM cannot ensure demand, but it can make earning from a small fragment of time, experience, or ability more possible than existing institutions do.

\subsection{One-way broadcasting vs. limited reverse access}\label{one-way-broadcasting-vs-limited-reverse-access}

Executives, authors, creators, and public figures can address large audiences, while individuals have little access in return. A chief executive could offer limited customer-call time and raise the price until demand becomes manageable. Customers with a strong need could connect directly, and the executive could hear information that organizational layers filter out. Payment purchases access, not favorable treatment.

\section{Human-Value Categories and Examples}\label{human-value-categories-and-examples}

BM does not require claims that AI can never perform a particular activity. Human participation remains valuable when the buyer cares about the source, relationship, identity, authority, or physical and social reality of the interaction. Seven categories illustrate this value.

\textbf{Firsthand experience and participant knowledge.} A student, resident, worker, caregiver, customer, former patient, or person who has undergone a difficult transition can explain participation from the inside. Its value lies in relevant lived experience, subject to ordinary limits of accuracy, privacy, and confidentiality.

\textbf{Human presence, conversation, and emotional support.} Sometimes the service is that another person is genuinely there---listening, responding, sharing time, offering encouragement, or helping someone feel less alone. A buyer may simply want a human conversation, including to pass time.

\textbf{Identity-specific access.} A buyer may want one particular executive, author, creator, witness, owner, or community member. Similar information from a substitute does not provide the same value because identity is part of the request.

\textbf{Current observation.} A person can see, hear, handle, or enter something the buyer cannot access. A resident might show an apartment at night, an owner might demonstrate a used object, or a local participant might report conditions at an unfolding event.

\textbf{Authority, consent, and accountability.} Some interactions matter because a person can decide, authorize, negotiate, consent, or accept professional responsibility. The institutional standing of the actual stakeholder is part of the service.

\textbf{Ordinary-person participation and feedback.} A designer may want a five-minute reaction from someone in a target group; a researcher may need a participant rather than an expert. BM lowers the cost of recruiting and compensating that participation.

\textbf{Personalized live performance.} A person may sing a song, tell a story or joke, conduct a role-play, or demonstrate a small creative ability in response to a buyer. The service combines live adaptation with that person\textquotesingle s participation.

These categories explain why \emph{helper} is more inclusive than \emph{expert}: BM includes expertise without being confined to it.

\section{Long-Run Predictions, Risks, and Research}\label{long-run-predictions-risks-and-research}

The following predictions depend on sufficient demand, market thickness, trust, and responsible governance.

\subsection{Workforce shift and new services}\label{workforce-shift-and-new-services}

If BM becomes sufficiently liquid, some human contribution could shift from conventional employment, formal engagements, public content production, and unpaid informal help into paid burst exchange. Many people would combine BM with employment, study, self-employment, care, or retirement. Some calls would lead to larger conventional engagements, including professional contracts or employment.

BM could support earning opportunities that cannot be inferred from present occupations. YouTube\textquotesingle s lower participation and distribution barriers helped ordinary people create unforeseen content and work. BM may similarly enable new live services while requiring less preparation and post-production. The comparison establishes a plausible innovation mechanism, not a forecast of equal scale.

\subsection{Professional prices and access}\label{professional-prices-and-access}

BM can reduce office, acquisition, scheduling, packaging, and billing costs for professional services suitable for brief video delivery. Evidence from unbundled legal service suggests that purchasing limited parts of a matter can reduce total cost and widen access (Solicitors Regulation Authority, 2023). Making suitable preliminary time independently billable could further reduce the need to recover intake costs elsewhere.

\begin{bmproposition}{Proposition 7 --- BM can reduce suitable professional-service prices}
For professional services suitable for brief video delivery, lower office, acquisition, scheduling, packaging, and uncompensated-intake costs should reduce the minimum purchase and effective price of obtaining help. Effects on the time rate depend on competition, platform fees, preparation, professional obligations, and demand.
\end{bmproposition}

The strongest prediction concerns the minimum purchase and total cost of resolving a narrow need; the time rate may also fall. In suitable online medical care, a patient could reach a doctor directly, while time-metered payment compensates the doctor for initial information collection. Evidence on gatekeeping and telemedicine continuity shows why clinical triage, history, records, licensing, examination, and emergency escalation remain necessary where applicable (Sripa et al., 2019; Graetz et al., 2024).

\subsection{Poverty, aging, productivity, and industrial change}\label{poverty-aging-productivity-and-industrial-change}

Productive work is an important route out of poverty, although many people remain poor in low-paid or informal work (International Labour Organization, 2016, 2026). If BM creates demand for capabilities held by lower-income participants, it could supplement income and reduce some public support burden. Lifelong Working could likewise give some older people voluntary income, purpose, connection, and a way to share accumulated experience.

The larger effect may be cumulative. Many small exchanges could accelerate learning, improve decisions, reveal needs, connect collaborators, generate product feedback, and create services and businesses. These ripples could contribute to productivity, prosperity, and industrial change even when each transaction is modest. Such effects should be measured rather than assumed.

BM may intensify competition with conventional services while supplying leads for larger engagements. One person has limited time, so growing demand can spill toward substitutes when suitable alternatives exist; identity-specific demand and platform concentration can limit this effect.

\subsection{Risks and research agenda}\label{risks-and-research-agenda}

Low entry and rapid connection increase the importance of identity, trust, privacy, and safety. Ratings can be manipulated or repeatedly direct demand toward established providers. Personal-access markets create risks of harassment, discrimination, exploitation, and emotional harm. Time billing can encourage delay or rushing, and providers may spend unpaid time waiting, preparing, and maintaining visibility. Higher-risk services require proportionate safeguards and professional responsibility.

Network effects, rankings, proprietary data, accumulated reputation, fees, and switching costs can concentrate intermediary power (Vallas and Schor, 2020). Responsible design may require transparent rankings and fees, appeals, data and reputation portability, accessibility, payment protection, and safeguards proportionate to service risk.

Research should begin with marketplace experiments. Studies can vary search effort, connection delay, pricing control, billing granularity, and video availability to test the minimum viable exchange. Pilots should measure transaction initiation, abandonment, duration, price, waiting and preparation time, repeat use, provider diversity, disputes, harm, and earnings after fees. Longitudinal studies should test whether new services emerge, work shifts, demand reaches underused providers, and suitable professional services become more accessible. BM\textquotesingle s employment, poverty, productivity, and welfare effects remain open empirical questions.

\section{Conclusion}\label{conclusion}

Much human capability remains unexchanged because the value of a brief contribution can be smaller than the cost of arranging its exchange. BM responds with a market for brief, paid, immediate human services delivered through live video. In the AI era, this longstanding utilization gap has become an urgent employment question as well as an economic opportunity.

Its economic mechanism is a lower minimum viable exchange. The four-part friction map covers buyer access, provider market access, capability packaging, and assurance and governance. Interactive elicitation lets useful capability emerge without a polished one-way product. PIVOT combines Provider-set price, Immediate access, Video interaction, Open participation, and Time-metered payment.

BM advances seven rebalances and serves both scarce and underused capacity. As generative market infrastructure, it may do more than shorten familiar services: it may expose overlooked human value, support a workforce shift, reduce suitable professional-service prices, improve collaboration, and generate services that cannot yet be foreseen.

These claims require evidence and responsible governance. If its effects emerge at scale, BM could represent the next leap for humanity.

\section*{Acknowledgments}

The author thanks Professor Marcel Ausloos for his insights, critiques, and counsel, and Dr. Bo Hu for his advice and support.

\section*{References}
\begingroup
\small
\setlength{\parindent}{-1.4em}
\setlength{\leftskip}{1.4em}
\setlength{\parskip}{0.55em}

Acemoglu, Daron, and Pascual Restrepo. 2019. ``Automation and New Tasks: How Technology Displaces and Reinstates Labor.'' \emph{Journal of Economic Perspectives} 33(2): 3--30. \url{https://doi.org/10.1257/jep.33.2.3}

Acquatella, François, Valérie Fernandez, and Thomas Houy. 2022. ``A Technical-Economic Perspective on Artificial Intelligence Role in Platform Markets Organization.'' \emph{Systèmes d\textquotesingle Information et Management} 27(4): 51--73. \url{https://doi.org/10.3917/sim.224.0051}

ARCEP/IDATE. 2006. \emph{Étude sur les acteurs du logiciel de communication}. \url{https://www.arcep.fr/uploads/tx_gspublication/etude-idate-acteurs-logiciel-oct06.pdf}

Bakos, J. Yannis. 1997. ``Reducing Buyer Search Costs: Implications for Electronic Marketplaces.'' \emph{Management Science} 43(12): 1676--1692. \url{https://doi.org/10.1287/mnsc.43.12.1676}

Becker, Gary S. 1965. ``A Theory of the Allocation of Time.'' \emph{Economic Journal} 75(299): 493--517. \url{https://doi.org/10.2307/2228949}

Brynjolfsson, Erik, Danielle Li, and Lindsey R. Raymond. 2025. ``Generative AI at Work.'' \emph{Quarterly Journal of Economics} 140(2): 889--942. \url{https://doi.org/10.1093/qje/qjae044}

Burgess, Jean, and Joshua Green. 2018. \emph{YouTube: Online Video and Participatory Culture}. 2nd ed. Cambridge: Polity.

Coase, Ronald H. 1937. ``The Nature of the Firm.'' \emph{Economica} 4(16): 386--405. \url{https://doi.org/10.1111/j.1468-0335.1937.tb00002.x}

Cowan, Robin. 2001. ``Expert Systems: Aspects of and Limitations to the Codifiability of Knowledge.'' \emph{Research Policy} 30(9): 1355--1372. \url{https://doi.org/10.1016/S0048-7333(01)00156-1}

Contzen, Sandra, Karin Zbinden, Cécile Neuenschwander, and Michèle Métrailler. 2017. ``Retirement as a Discrete Life-Stage of Farming Men and Women\textquotesingle s Biography?'' \emph{Sociologia Ruralis} 57(S1): 730--751. \url{https://doi.org/10.1111/soru.12154}

Cunningham, Stuart, David Craig, and Jon Silver. 2016. ``YouTube, Multichannel Networks and the Accelerated Evolution of the New Screen Ecology.'' \emph{Convergence} 22(4): 376--391. \url{https://doi.org/10.1177/1354856516641620}

Dahlman, Carl J. 1979. ``The Problem of Externality.'' \emph{Journal of Law and Economics} 22(1): 141--162. \url{https://doi.org/10.1086/466936}

Felső, Flóra, Sander Onderstal, and Jo Seldeslachts. 2022. ``The Pricing Structure of Legal Services: Do Lawyers Offer What Clients Want?'' \emph{Review of Industrial Organization} 61(2): 123--148. \url{https://doi.org/10.1007/s11151-022-09868-9}

Gmyrek, Pawel, Janine Berg, Karol Kamiński, Filip Konopczyński, Agnieszka Ładna, Balint Nafradi, Konrad Rosłaniec, and Marek Troszyński. 2025. \emph{Generative AI and Jobs: A Refined Global Index of Occupational Exposure}. ILO Working Paper 140. Geneva: International Labour Organization. \url{https://doi.org/10.54394/HETP0387}

Gonzalez, Nick. 2006. ``BitWine Gives Access to Those in the Know.'' \emph{TechCrunch}, November 28. \url{https://techcrunch.com/2006/11/28/bitwine-gives-acces-to-those-in-the-know/}

Gosain, Sanjay. 2007. ``Mobilizing Software Expertise in Personal Knowledge Exchanges.'' \emph{Journal of Strategic Information Systems} 16(3): 254--277. \url{https://doi.org/10.1016/j.jsis.2007.02.001}

Graetz, Ilana, Jie Huang, Anjali Gopalan, Emilie Muelly, Andrea Millman, and Mary E. Reed. 2024. ``Primary Care Telemedicine and Care Continuity: Implications for Timeliness and Short-Term Follow-Up Healthcare.'' \emph{Journal of General Internal Medicine} 39(13): 2454--2460. \url{https://doi.org/10.1007/s11606-024-08914-4}

Hadfield, Gillian K. 2019. ``More Markets, More Justice.'' \emph{Dædalus} 148(1): 37--48. \url{https://doi.org/10.1162/DAED_a_00533}

International Labour Organization. 2016. \emph{World Employment and Social Outlook 2016: Transforming Jobs to End Poverty}. Geneva: ILO. \url{https://www.ilo.org/publications/world-employment-and-social-outlook-2016-transforming-jobs-end-poverty}

International Labour Organization. 2026. \emph{Employment and Social Trends 2026}. Geneva: ILO. \url{https://www.ilo.org/publications/flagship-reports/employment-and-social-trends-2026}

Jaakkola, Elina. 2020. ``Designing Conceptual Articles: Four Approaches.'' \emph{AMS Review} 10: 18--26. \url{https://doi.org/10.1007/s13162-020-00161-0}

Jacob, Karl, Scott Faber, and Sean Van der Linden. 2007. ``Method and System to Connect Consumers to Information.'' U.S. Patent 7,224,781, issued May 29. \url{https://patents.google.com/patent/US7224781B2/en}

Kadane, Joseph B., James M. Dickey, Robert L. Winkler, Wayne S. Smith, and Stephen C. Peters. 1980. ``Interactive Elicitation of Opinion for a Normal Linear Model.'' \emph{Journal of the American Statistical Association} 75(372): 845--854. \url{https://doi.org/10.1080/01621459.1980.10477562}

Keen. n.d.-a. ``Find a Psychic: Best Psychics.'' Accessed September 13, 2026. \url{https://www.keen.com/}

Keen. n.d.-b. ``Getting Started on Keen: Your First Reading and What to Expect.'' \emph{Keen Help Center}. Accessed September 13, 2026. \url{https://help.keen.com/hc/en-us/articles/50919969977747-Getting-Started-on-Keen-Your-First-Reading-What-to-Expect}

Li, Jun, and Serguei Netessine. 2020. ``Higher Market Thickness Reduces Matching Rate in Online Platforms: Evidence from a Quasiexperiment.'' \emph{Management Science} 66(1): 271--289. \url{https://doi.org/10.1287/mnsc.2018.3223}

Lomas, Natasha. 2021. ``Anyone Is Building a Marketplace for Advice, One 5-Minute Call at a Time.'' \emph{TechCrunch}, July 14. \url{https://techcrunch.com/2021/07/14/anyone-is-building-a-marketplace-for-advice-one-5-minute-call-at-a-time/}

Malone, Thomas W., JoAnne Yates, and Robert I. Benjamin. 1987. ``Electronic Markets and Electronic Hierarchies.'' \emph{Communications of the ACM} 30(6): 484--497. \url{https://doi.org/10.1145/214762.214766}

Manber, Udi. 2013. ``Introducing Helpouts: Help When You Need It over Live Video.'' \emph{Google Blog}, November 4. \url{https://blog.google/products-and-platforms/products/google-plus/introducing-helpouts-help-when-you-need/}

Microsoft. 2001. ``MSN Teams with Keen to Deliver Live, Immediate Advice across MSN.'' July 27. \url{https://news.microsoft.com/source/2001/07/27/msn-teams-with-keen-to-deliver-live-immediate-advice-across-msn/}

Oswald, Ed. 2006. ``Ether Looking to Be `eBay of Services.'\,'' \emph{BetaNews}, March 1. \url{https://betanews.com/article/ether-looking-to-be-ebay-of-services/}

Ribeiro, John. 2013. ``Google Introduces Helpouts, a New Way to Get Help over Video.'' \emph{PCWorld}, November 5. \url{https://www.pcworld.com/article/448470/google-introduces-helpouts-a-new-way-to-get-help-over-video.html}

Roth, Alvin E. 2008. ``What Have We Learned from Market Design?'' \emph{Economic Journal} 118(527): 285--310. \url{https://doi.org/10.1111/j.1468-0297.2007.02121.x}

Rubin, Matan, Joanna Z. Li, Federico Zimmerman, Desmond C. Ong, Amit Goldenberg, and Anat Perry. 2025. ``Comparing the Value of Perceived Human versus AI-Generated Empathy.'' \emph{Nature Human Behaviour} 9(11): 2345--2359. \url{https://doi.org/10.1038/s41562-025-02247-w}

Rysman, Marc. 2009. ``The Economics of Two-Sided Markets.'' \emph{Journal of Economic Perspectives} 23(3): 125--143. \url{https://doi.org/10.1257/jep.23.3.125}

Solicitors Regulation Authority. 2023. ``Unbundled Services Pilot: Final Report.'' June 15. \url{https://www.sra.org.uk/unbundled-services-pilot}

Sripa, Poompong, Benedict Hayhoe, Priya Garg, Azeem Majeed, and Geva Greenfield. 2019. ``Impact of GP Gatekeeping on Quality of Care, and Health Outcomes, Use, and Expenditure: A Systematic Review.'' \emph{British Journal of General Practice} 69(682): e294--e303. \url{https://doi.org/10.3399/bjgp19X702209}

ten Raa, Thijs, and Victoria Shestalova. 2004. ``Empirical Evidence on Payment Media Costs and Switch Points.'' \emph{Journal of Banking \& Finance} 28(1): 203--213. \url{https://doi.org/10.1016/S0378-4266(02)00404-1}

Vallas, Steven P., and Juliet B. Schor. 2020. ``What Do Platforms Do? Understanding the Gig Economy.'' \emph{Annual Review of Sociology} 46: 273--294. \url{https://doi.org/10.1146/annurev-soc-121919-054857}

von Hippel, Eric. 2005. \emph{Democratizing Innovation}. Cambridge, MA: MIT Press. \url{https://doi.org/10.7551/mitpress/2333.001.0001}

Wenger, Joshua D., C. Daryl Cameron, and Michael Inzlicht. 2026. ``People Choose to Receive Human Empathy Despite Rating AI Empathy Higher.'' \emph{Communications Psychology} 4: Article 19. \url{https://doi.org/10.1038/s44271-025-00387-3}

Williamson, Oliver E. 1981. ``The Economics of Organization: The Transaction Cost Approach.'' \emph{American Journal of Sociology} 87(3): 548--577. \url{https://doi.org/10.1086/227496}

Zha, Vincent. 2021. ``Burst Market---the Next Leap of Humanity.'' arXiv:2112.06646v1. \url{https://arxiv.org/abs/2112.06646v1}
\par
\endgroup

\end{document}